\documentclass[10pt,twocolumn,twoside]{IEEEtran}

\usepackage{amsmath}
\usepackage{bbold}
\usepackage{enumerate}
\usepackage{cite}
\usepackage{color,xcolor}
\usepackage{amssymb}
\usepackage{graphicx}
\usepackage{algorithm}
\usepackage{algpseudocode}
\usepackage{bbm}
\usepackage{graphics}
\usepackage{subfigure}
\usepackage{epsfig}
\usepackage{epstopdf}
\usepackage{makecell}

\begin{document}
		\newtheorem{lemma}{Lemma}
	\newtheorem{assumption}{Assumption}
	\newtheorem{theorem}{Theorem}
	\newtheorem{corollary}{Corollary}
	\newtheorem{remark}{Remark}
	\newtheorem{problem}{Problem}
	\newenvironment{proof}{\hspace{0ex}\textsc{Proof}.\hspace{1ex}}{\hfill$\Box$\newline}
\title{Harmonic-Coupled Riccati Equation and its Applications in Distributed Filtering}

\author{Jiachen Qian, Peihu Duan, Zhisheng Duan*, \IEEEmembership{Senior member, IEEE}, Ling Shi, \IEEEmembership{Fellow, IEEE}
	\thanks{This work is supported by the National Natural Science Foundation of China under Grants T2121002 and 62173006, the Hong Kong
		Research Grants Council under the General Research Fund Grant 16211622. (*$Corresponding\;author: Zhisheng\;Duan)$}
	\thanks{J. Qian and Z. Duan are with State Key Laboratory for Turbulence and Complex Systems, Department of Mechanics and Engineering Science, College of
		Engineering, Peking University, Beijing, 100871, China. P. Duan is with School of Electrical Engineering and Computer Science, KTH Royal Institute of Technology, Stockholm, Sweden. 
		L. Shi is with the Department of Electronic and Computer Engineering, the Hong Kong University of Science and Technology, Clear Water Bay, Kowloon,
		Hong Kong Special Administrative Region of China. E-mails: duanzs@pku.edu.cn (Z. Duan), jcq@pku.edu.cn (J. Qian), peihu@kth.se (P. Duan), eesling@ust.hk (L. Shi).}}

\maketitle

\begin{abstract}
    The coupled Riccati equations (CREs) are a set of multiple Riccati-like equations whose solutions are coupled with each other through matrix means. They are a fundamental mathematical tool to depict the inherent dynamics of many complex systems, including Markovian systems or multi-agent systems.
    This paper investigates a new kind of CREs called harmonic-coupled Riccati equations (HCREs), whose solutions are coupled using harmonic means.
    We first introduce the specific form of HCREs and then analyze the existence and uniqueness of its solutions under the conditions of collective observability and primitiveness of coupling matrices. Additionally, we ensure the uniqueness of HCREs solutions with several mild conditions.
    Based on this newly established theory, we greatly simplify the steady-state estimation error covariance of consensus-on-information-based distributed filtering (CIDF)
     into the solutions to a discrete-time Lyapunov equation (DLE). This leads to a significant conservativeness reduction of traditional performance evaluation techniques for CIDF.
    The obtained results are remarkable since they not only enrich the theory of CREs, but also provide a novel insight into the synthesis and analysis of CIDF algorithms. We finally validate our theoretical findings through several numerical experiments.
\end{abstract}


\section{Introduction}
\subsection{Background}
The Riccati equation, particularly the algebraic Riccati equation, provides a solid theoretical foundation for control and filtering technologies. For instance, it can be used for performance evaluation of Kalman filter \cite{Andersonoptimal}, linear quadratic regulator (LQR) design\cite{willems1971least}, and behavior assessment in dynamic noncooperative games \cite{bacsar1998dynamic}. Hence, a comprehensive investigation of the Riccati equation can greatly advance the development of related fields.
 
With the emergence of networked control systems, research object has shifted from single linear time-invariant systems to more complex systems, such as Markovian systems and multi agent systems. In these scenarios, traditional theories about a single Riccati equation are insufficient to capture system properties, such as correlations induced by the random jumps of the system and information flow within the network.
As a result, there is an urgent need to develop more effective and general mathematical tools, while coupled Riccati equations (CREs) are one of the most critical techniques among them.

\subsection{Coupled Riccati Equations}
The CREs consist of multiple Riccati-like equations, with their solutions coupled with each other in the form of matrix means such as algebraic mean. These equations were first formulated from the Markovian-jump LQR problem \cite{blair1975feedback, chizeck1986discrete} and optimal controllers can be designed using solutions to the corresponding algebraic coupled Riccati-like equations (ACREs), i.e.,
\begin{equation}\label{oriACRE}
\begin{aligned}
&P_{i} = A\Big(\sum_{j=1}^{N}l_{ij}\tilde{P}_{j}^{-1}+l_{ij}C_j^TR_j^{-1}C_j\Big)^{-1}A^{T} + Q,\\
&\tilde{P}_i\triangleq\sum_{j=1}^{N}l_{ij}P_{j},\;
\end{aligned}
\end{equation}
where $i,j=1,\;2,\;\cdots,\;N$, $A$ and $C\triangleq\big[C_1^T,\,C_2^T,\,\cdots,\, C_N^T\big]^T$ are system matrices, $l_{ij}$ is the $(i,j)$-th element of any stochastic matrix $\mathcal{L}$, $Q$ and $R\triangleq \text{diag}\big\{R_1,\dots,R_N\big\}$ are positive definite, and $P_i$ is the soloution to each Riccati-like equation coupled with each other in the form of algebraic means $\tilde{P}_i$. The basic theories of ACREs, including the existence of solutions to ACREs \cite{chizeck1986discrete, abou1994solution, abou1995solution, costa1999maximal}, upper and lower matrix bounds of the solutions \cite{davies2008upper, czornik2001upper, lee2006new}, and numerical algorithms to obtain the solutions \cite{abou1994solution, li2022inversion}, have been extensively investigated in the literature. 

However, in some scenarios, such as information-weighted distributed state estimation \cite{kamal2013information, battistelli2014kullback, battistelli20156897960, he2018consistent, wang2017convergence}, the harmonic matrix mean rather than the algebraic one is primarily utilized for node interaction. Therefore, the theories of ACREs are infeasible here. As a remedy, another kind of CREs that involve harmonic means are introduced to demonstrate the properties of these situations. Compared with ACREs proposed in \cite{chizeck1986discrete, abou1994solution, abou1995solution}, the coupled term $\tilde{P}_i$ in HCREs is converted from algebraic mean to harmonic mean, i.e.,
$$
\tilde{P}_i\triangleq\sum_{j=1}^{N}l_{ij}P_{j}\;\;\Rightarrow\;\;\tilde{P}_i\triangleq\Big(\sum_{j=1}^{N}l_{ij}P_{j}^{-1}\Big)^{-1},
$$
which adopts a nonlinear form and embodies stronger couplings between solutions $P_i,\;i=1,\cdots,N$. 

The existing results on HCREs primarily focus on two control areas: consensus-on-information-based distributed filtering (CIDF) \cite{battistelli2014kullback, battistelli20156897960, battistelli2018distributed, he2018consistent, he2020distributed, duan2019distributed} and LQR cooperative regulator \cite{duan2020distributed}. In the CIDF framework, recent studies have shown that harmonic means, also referred to as ``covariance intersection fusion" in \cite{battistelli2014kullback}, can guarantee the stability of the distributed filter with necessary and sufficient requirements on network connectivity, system observability and fusion steps \cite{battistelli2014kullback, battistelli20156897960, he2018consistent, duan2020distributed}.  Specifically, Battistelli {\it et al.} \cite{battistelli2014kullback} formulated a CIDF framework by embedding the matrix harmonic mean into the traditional Kalman filter. 
Specifically, the matrix iteration of CIDF, i.e.,
\begin{equation}\label{MILCIDF}
P_{i,k+1|k} = A\Big(\sum_{j=1}^{N}l_{ij}P_{j,k|k-1}^{-1}+l_{ij}C_j^TR_j^{-1}C_j\Big)^{-1}A^{T} + Q,
\end{equation}
is shown to be bounded under a mild collective observability of $\left(A,C\right)$ and primitivity of the weighting matrix $\mathcal{L}$. 
In this setting, HCREs can be interpreted as the steady-state form of the matrix iterative law \eqref{MILCIDF}.
Some studies have investigated the properties of the iteration \eqref{MILCIDF} or its variants to evaluate the performance of CIDF frameworks. 
For instance, Battistelli {\it et al.} \cite{battistelli20156897960} proposed a hybrid information fusion framework, which adopted two weighting matrices for the information terms $P_{j,k|k-1}$ and $C_j^TR_j^{-1}C_j$ to improve the performance of CIDF.
Wang {\it et al.} \cite{wang2017convergence} studied the iteration \eqref{MILCIDF} with time-varying weighting matrix $\mathcal{L}_k$ and relaxed the requirement of the weighting matrix to joint connectivity.
He {\it et al.} \cite{he2018consistent} discussed the stability of CIDF algorithms with a more general system model \cite{he2018consistent}, where state matrix $A$ was non-invertible for some time steps. 
Duan {\it et al.} \cite{duan2020distributed} generalized the matrix iteration \eqref{MILCIDF} for the perturbation of matrix $A$ and demonstrated the convergence of $P_{i,k+1|k}$ with some elaborated but strict initial conditions.

However, all above results on CIDF related to HCREs are subjected to one fundemental problem, i.e., 
the stability of the iteration \eqref{MILCIDF} can only be guaranteed by proving the boundedness of $P_{i,k+1|k}$ with huge conservativeness \cite{battistelli2014kullback, battistelli20156897960, he2018consistent, wang2017convergence}.
Even though it was shown in \cite{duan2020distributed} that $P_{i,k+1|k}$ \eqref{MILCIDF} converges to a fixed point for some specific initial conditions $P_{1|0}$, the hypothesis was too strict to satisfy in practice.  
Generally speaking, the upper bound of $P_{i,k+1|k}$ derived in the literature is much restrictive and more precise properties of CIDF, such as the convergence of the matrix iterative law and the steady-state performance, remain not fully found. The in-depth discussion on the conservatism of the previous studies on CIDF is lacking.  

From above discussion, it is beneficial for understanding the performance of CIDF, particularly its steady-state behavior, if we have a good understanding of HCREs. 
However, this is not a handy task. Firstly, the harmonic mean $\tilde{P}_i\triangleq\big(\sum_{j=1}^{N}l_{ij}P_{j}^{-1}\big)^{-1}$ in the Riccati equations makes the correlation between $P_i$ extremely nonlinear and non-convex. Secondly, instead of each local $\big(A,C_i\big)$, the collective $\big(A,C\big)$ is assumed to be observable, which means that traditional mathematical techniques used for analyzing algebraic Riccati equation \cite{kailath2000linear} and ACREs \cite{chizeck1986discrete} are no longer applicable. Therefore, new mathematical techniques are urgently needed to  excavate more properties of HCREs.
The ongoing research is likely to enhance our understanding of these equations and their solutions.
 
\subsection{Contributions}
Building on previous discussions, this paper is aimed to uncover more valuable properties of HCREs. Specifically, we attempt to establish some sufficient conditions for the existence and uniqueness of solutions to HCREs and develop novel mathematical techniques for obtaining these solutions. By leveraging these outcomes, we aim to derive a closed-form expression for the steady-state performance of CIDF and construct a systematical performance evaluation framework.  
The contributions of this paper are summarized as follows:
\begin{enumerate}
	\item  
	Only two mild requirements, namely the collective observability of $\big(A,C\big)$ and the primitivity of the information weighting matrix $\mathcal{L}$, are required to ensure the existence and uniqueness of solutions to HCREs ({\bf Theorem \ref{thmunique}}). Fundamental mathematical techniques for the analysis of HCREs are developed. These new findings greatly enrich the HCREs theory.
	\item 
	\textcolor{black}{In addition to the basic theory of HCREs, we manage to find a low computation-complexity iterative law to obtain the solutions to HCREs. It is demonstrated that the matrix iterative law of CIDF guarantees convergence of $P_{i,k+1|k}$ to the solution $P_i$ of the HCREs, regardless of initial value ({\bf Theorem \ref{thmconverge}}).} This result provides a constructive insight into the steady-state behavior of the CIDF matrix iterative law and further reveals the essence of the stability of CIDF.

	\item By applying the obtained novel theories of HCREs, it has been demonstrated that the estimation error covariance matrix of CIDF converges. The closed-form of the steady-state covariance matrix can be simplified as the solution to a discrete-time Lyapunov equation (DLE) ({\bf Section \ref{seccovariance}}). Furthermore, the proposed HCREs framework unifies all classical CI-based distributed filtering algorithms \cite{kamal2013information, battistelli2014kullback, battistelli20156897960}, only differing in the parameter matrix $\mathcal{L}$. These precise results are established for the CIDF for the first time.  
\end{enumerate} 

The remainder of this paper is organized as follows. {Some preliminaries}, including the background of HCREs and the problem formulation, are presented in Section \ref{sec2}. The main results, including the analysis of the solution to HCREs and the application of HCREs to CIDF, are presented in Sections \ref{sec3} and \ref{secdiscussion}. Some illustrative numerical experiments are presented in Section \ref{secsimulation}. Conclusions are {drawn} in Section \ref{secconcu}.

\textit{Notation:} For two symmetric matrices $X_{1}$ and $X_{2}$, $X_{1}> X_{2}\left(X_{1}\ge X_{2}\right)$ means that $X_{1}-X_{2}$ is positive definite (positive semi-definite). {$\text{exp}\left(\cdot\right)$} denotes the exponential function. $\big|a\big|$ denotes the absolute value of real number $a$ or the norm of complex number $a$.  $\mathcal{L}\rhd 0\left(\unrhd 0\right)$ means that all the elements of matrix $\mathcal{L}$ are positive (non-negative). $\mathbb{E}\left\lbrace x\right\rbrace$ denotes the expectation of a random variable $x$. $\lambda\left(A\right)$ denotes the eigenvalue of matrix $A$. $\rho\left(A\right)$ denotes the spectral radius of $A$.  $\left\|A\right\|_2$ denotes the 2-norm (the largest singular value) of matrix $A$. $X\otimes Y$ denotes the Kronecker product of matrix $X$ and $Y$. $I_n$ denotes the identity matrix with dimention $n$. 

\section{Preliminaries and Problem Formulation}\label{sec2}
In this section, the system model and the CI-based distributed filtering algorithm are provided. In addition, a comprehensive literature review of CIDF and the problem formulation of HCREs are also given.
\subsection{{System} Model}\label{subModel}
{Consider} a network of $N$ sensors that measure and estimate the states of a linear time-invariant system, described by
\begin{equation}
\begin{aligned}
&x_{k+1}=A x_k + \omega_k,\quad k = 0,1,2,\ldots,\\
&y_{i,k}=C_{i}x_k + v_{i,k},\quad i = 1,2,\ldots,N,
\end{aligned}
\end{equation}
where $x_k\in\mathbb{R}^n$ is the state vector of the system, $y_{i,k}\in\mathbb{R}^{m_i}$ is the measurement vector of sensor $i$,  $\omega_k\in\mathbb{R}^n$ is the process noise with covariance $Q>0\in\mathbb{R}^{n\times n}$, and $v_{i,k}\in\mathbb{R}^{m_i}$ is the observation noise with covariance $R_{i}>0\in\mathbb{R}^{m_i \times m_i}$. The sequences $\left\{\omega_k\right\}^{\infty}_{k=0}$ and $\left\{v_{i,k}\right\}^{\infty,N}_{k=0,i=1}$ are assumed to be \textcolor{black}{mutually uncorrelated white Gaussian noise}. Besides, $A$ is the state-transition matrix and $C_{i}$ is the observation matrix of sensor $i$. Moreover, let $C=\left[C_{1}^T,C_{2}^T,\dots,C_{N}^T\right]^T$ and $R=\text{diag}\left\{R_{1},\dots,R_{N} \right\}$. 

The communication topology of the sensor network is denoted by $\mathcal{G}=\left(\mathcal{V},\mathcal{E},\mathcal{L}\right)$, where $\mathcal{V}=\left\lbrace1,2,\dots,N\right\rbrace$ is the node set, $\mathcal{E}\subseteq \mathcal{V}\times\mathcal{V}$ is the edge set, and $\mathcal{L}=\left[l_{ij}\right]$ is the adjacency matrix. The adjacency matrix reflects the interactions among the nodes, e.g., $l_{ij}>0\Leftrightarrow\left(i,j\right)\in\mathcal{E}$, which means that sensor $i$ can receive information from sensor $j$. In this case, sensor $j$ is called an in-neighbor of sensor $i$, and sensor $i$ is called an out-neighbor of sensor $j$. {For simplicity, let $i\in\mathcal{V}$ represent the $i$-th sensor of the network.} $\mathcal{N}_{i}$ denotes the in-neighbor set of sensor $i$, and $l_{i}$ denotes the $i$-th row of $\mathcal{L}$.
$l_{ij}^{(L)}$ is the $(i,j)$-th element of matrix $\mathcal{L}^L$.
%
In this paper, let $\bar{\mathcal{L}}$ denote a special kind of adjacency matrix, where the $\left(i,j\right)$-th element of $\bar{\mathcal{L}} \ \text{is}\  1\Leftrightarrow\left(i,j\right)\in\mathcal{E}$. 
{The diameter $d$ of graph $\mathcal{G}$ is the length of the longest path between any two nodes in the graph.}

\subsection{Information Weighted Distributed Filtering}\label{secCIDF}
In this subsection, some typical CIDF algorithms are introduced, which is the background of HCREs and one important motivation. 

Different from Kalman filter \cite{kalmanfilter},  CIDF consists of two main parts.
The first part is a local Kalman filter, where each sensor performs the Kalman iterative law to obtain the {\it a posterior} state estimates using local observations. 
The second part is the covariance intersection-based information fusion step, where each sensor combines the {\it a priori} information received from neighbouring nodes to compute a more precise estimate.

The stability of CIDF under weak observability is usually ensured by the covariance intersection-based fusion technique. This technique was first proposed in \cite{julier1997non, niehsen2002information}, where the matrix intersection technique was used to guarantee the consistency of the estimator when there are unknown information correlations. In \cite{hu2011diffusion}, Hu {\it et al.} applied the covariance intersection technique to solve the distributed estimation problem and formulated the diffusion Kalman filter. However, this approach required a large number of fusion steps among sensors to achieve the local observability to further guarantee the filter stability. To address this issue, Battistelli {\it et al.} \cite{battistelli2014kullback} formulated the CIDF framework that achieved the filter stability with a weak observability and a small number of fusion steps between two sampling instants. Since then, extensive efforts have been devoted to improving the performance of CIDF algorithms with hybrid fusion structures \cite{battistelli20156897960}, reducing energy costs through event-based communication mechanisms \cite{battistelli2018distributed}, ensuring the stability of the CIDF algorithm with a more general system model \cite{he2018consistent}, and addressing model uncertainty with a modified CIDF algorithm \cite{duan2020distributed}.

While the theory of CI-based distributed filtering becomes sophisticated, several significant challenges remain unresolved. 
In the literature regarding CIDF algorithms \cite{battistelli2014kullback, battistelli20156897960, he2018consistent}, the stability of the designed filters was ensured through the proof of the uniform boundedness of the iterative term $P_{i,k}$ for all $i\in\mathcal{V}$ and $k\in\mathbb{N}$, regardless of the initial value $P_{i,0}$. The key techniques involve using the observation Gramian matrix to obtain a lower bound of $P_{i,k}^{-1}$ as demonstrated in \cite{battistelli2014kullback, he2018consistent}. Afterward, a Lyapunov function can be constructed using $P_{i,k}$ to prove the stability of the noise-free feedback system. \textcolor{black}{Although these bounds of $P_{i,k}^{-1}$ can be further used to prove the boundedness of the estimation error covariance matrix, it may be restrictive and much larger than the actual value of $P_{i,k}$.} Moreover, the boundedness of $P_{i,k}$ does not reveal a direct connection between the filtering structure $\mathcal{L}$ and the filtering performance. Consequently, it is impossible to optimize the filter's performance through the parameter tuning.


Based on the above analysis, it becomes apparent that the current theoretical basis for CIDF greatly hampers its further development. Hence, it is necessary to develop new theories to address this issue.  

\subsection{Problem Formulation}\label{subAlgorithm}
Generally, the CIDF algorithms take the form demonstrated in Algorithm \ref{alg1}, where the matrix terms $P_{i,k|k-1}$ play an important part in the design of feedback gain $K_{i,k}$ and information fusion gain $l_{ij}P_{i,k}P_{j,k|k}^{-1}$.  
Meanwhile, the iterative law of $P_{i,k+1|k}$ is equavilently rewritten as \eqref{MILCIDF}.

\begin{algorithm}[h]
	\caption{Distributed Information Fusion Algorithm}
	\label{alg1}
	\textbf{Input:}\\
	$\text{  }$
	$\hat{x}_{i,0}, P_{i,0},\quad i\in\mathcal{V},$\\
	\textbf{Prediction:}\\
	$\text{  }$
	$\hat{x}_{i,k|k-1}=A\hat{x}_{i,k-1}$,\\
	$\text{   }$
	$P_{i,k|k-1}=AP_{i,k-1}A^T+Q,$\\
	\textbf{Correction:}\\
	$\text{    }$
	$z_{i,k}=y_{i,k}-C_{i,k}\hat{x}_{i,k|k-1},$ \\
	$\text{    }$	
	$K_{i,k}=P_{i,k|k-1}C_{i}^T\left(C_{i}P_{i,k|k-1}C_{i}^T+R_{i}\right)^{-1},$
	\\
	$\text{    }$
	$\hat{x}_{i,k|k}=\hat{x}_{i,k|k-1}+K_{i,k}z_{i,k},$\\
	$\text{    }$
	$P_{i,k|k}=P_{i,k|k-1}-K_{i,k}C_{i}P_{i,k|k-1},$\\
	\textbf{Information Fusion:}\\
	$\text{    }$
	$P_{i,k}= \left(\sum_{j=1}^N l_{ij}P_{j,k|k}^{-1}\right)^{-1},$\\
	$\text{    }$
	$\hat{x}_{i,k}=P_{i,k}\left(\sum_{j=1}^{N}l_{ij}P_{j,k|k}^{-1}\hat{x}_{j,k|k}\right).$
\end{algorithm}


It is evident that HCREs represent the steady-state form of the iterative law \eqref{MILCIDF}. By studying HCREs, researchers can gain a deeper understanding of the iterative law and develop the theory of CI-based distributed filtering. This, in turn, can provide valuable insights for the design of CI-based filtering algorithms and parameters. In this paper, we formulate two fundamental problems related to the HCREs:
\begin{enumerate}
	\item How to establish a quantitative relation between matrices $A,C,\mathcal{L}$ and properties of HCREs,  particularly leveraging the connetion between the iterative law \eqref{MILCIDF} and the solution to HCREs?
	
	\item Based on the theory of HCREs, how to develop a unified close-form performance evaluation technique of all CI-based distributed filtering algorithms?  
\end{enumerate}

\section{Hamonic-Coupled Riccati Equations}\label{sec3}
In this section, the basic properties of the solution to HCREs will be revealed, especially the existence and uniqueness of the solution, together with the iterative law to obtain the unique solution. To do so, two 
preliminary assumptions are presented here:
\begin{assumption}\label{assystem}
	The matrix $A$ is invertible and $\big(A,C\big)$ is observable.
\end{assumption}
\vspace{6pt}
\begin{assumption}\label{ascommunication}
	The matrix $\mathcal{L}$ is primitive and row stochastic, i.e., $\sum_{j=1}^{N}l_{ij}=1,\;\;\forall i\in\mathcal{V}$.
\end{assumption}
\vspace{6pt}

As proposed in \cite{battistelli2014kullback, battistelli20156897960, kamal2013information}, both of the above two assumptions
are mild for distributed filtering problems. Generally
speaking, the invertibility of $A$ in Assumption \ref{assystem} is automatically satisfied in sampled-data systems as the matrix $A$ is obtained through discretization of continuous-time systems. Meanwhile, the observability of $\big(A,C\big)$ is essential
for the stability of the filtering algorithm. As for Assumption 2, note that if the corresponding communicaton graph of $\mathcal{L}$ is strongly connected and the diagonal elements of $\mathcal{L}$ is positive, then the matrix $\mathcal{L}$ is primitive \cite{Horn1985}.
In this section, to simplyfy the notations, we replace the term $P_{i,k+1|k}$ by $P_{i,k}$ with slightly abuse of notations. Under Assumption \ref{assystem} and \ref{ascommunication},  we mainly aim to prove the following facts:
\begin{enumerate}
	\item {\textbf{(Uniqueness)}} The Hamonic-Coupled Riccati Equations, 
	\begin{equation}\label{HCRE}
	P_{i} = A\Big(\sum_{j=1}^{N}l_{ij}P_{j}^{-1}+l_{ij}C_j^TR_j^{-1}C_j\Big)^{-1}A^{T} + Q
	\end{equation}
	have one unique group of solutions.
	\item  \textbf{(Convergence)} The iterative law 
	\begin{equation}\label{HRiter}
	P_{i,k+1} = A\Big(\sum_{j=1}^{N}l_{ij}P_{j,k}^{-1}+l_{ij}C_j^TR_j^{-1}C_j\Big)^{-1}A^{T} + Q
	\end{equation}
 converges to the unique solution to the HCREs with $k\to\infty$, regardless of the initial value $P_{i,0}$.
\end{enumerate}

\subsection{Uniqueness of the Solution to HCREs}\label{Uniqueness}
In this subsection, the proof of the existence and uniqueness of the solution to the HCREs \eqref{HCRE} is first proposed.
{To do so, the following two Lemmas are needed}.
\vspace{6pt}
\begin{lemma}\label{GBlm}
	(\cite[Theorem 4]{battistelli2014kullback}) For any given matrices $A,C,Q,R,\mathcal{L}$ that satisfy Assumption \ref{assystem} and \ref{ascommunication}, there exist a number $\bar{k}$ and a matrix $P$, such that $P_{i,k}\leq P,\;\forall i\in\mathcal{V},k\ge\bar{k}$.
\end{lemma}
\vspace{6pt}
Similar to the idea proposed in Theorem 2 of \cite{duan2019distributed}, one has the following Lemma:
\vspace{6pt}
\begin{lemma}\label{Duanlm}
	Let Assumption \ref{assystem} and \ref{ascommunication} hold, and suppose the initial value of the iteration \eqref{HRiter} satisfies $P_{i,0}\leq\epsilon I,\;\forall i\in\mathcal{V}$. Then, for sufficient small $\epsilon$, there is $P_{i,k+1}\ge P_{i,k}$ and $P_{i,k}$ converges with the increase of $k$.
\end{lemma}
\vspace{6pt}
\begin{proof}
	The proof of this Lemma takes a similar argument to the proof of Theorem 2 in \cite{duan2020distributed}.
	
As $Q$ is a positive definite matrix, one can choose a sufficient small $\epsilon$, such that $P_{i,1}\ge P_{i,0},\;\forall i\in\mathcal{V}$. Then, for all $k\ge 1$, one has
$$
\begin{aligned}
&\big(P_{i,k+1}-Q\big)^{-1}-\big(P_{i,k}-Q\big)^{-1}\\
&\qquad\qquad=\big(A^{-1}\big)^T\Big(\sum_{j=1}^{N}l_{ij}\big(P_{j,k}^{-1}-P_{j,k-1}^{-1}\big)\Big)A^{-1}.
\end{aligned}
$$	
With mathematical induction, one has $P_{i,k+1}\ge P_{i,k},\;\forall i\in\mathcal{V}$. Together with the boundedness of $P_{i,k}$ proved in Lemma \ref{GBlm}, one can obtain that the iterative law \eqref{HRiter} converges with specific initial value $P_{i,0}$.
\end{proof}
\vspace{6pt}
\begin{lemma}\label{lmexistence}
	 For any given matrices $A,C$, $Q$, $R$, $\mathcal{L}$ that satisfy Assumption \ref{assystem} and \ref{ascommunication}, there exists at least one group of solution $\big\{P_i\big\}\triangleq\big\{P_1,P_2,\cdots,P_N\big\}$ to the hamonic-coupled Riccati equations \eqref{HCRE}.
\end{lemma}
\vspace{6pt}
\begin{proof}
    Denote the convergent value of $P_{i,k}$ in Lemma \ref{Duanlm} as $P_i$, then the result follows.
\end{proof}
\vspace{6pt}


Before the proof of the uniqueness of the solution to HCREs \eqref{HCRE}, the following notations are defined to simplify the proof.

For any $i\in\mathcal{V}$, the following notations are denoted as
$$
\tilde{C}_{i}=\Big[\text{sign}\big(l_{i1}\big) C_{1}^T,\cdots,\text{sign}\big(l_{iN}\big)C_{N}^T\Big]^T,
$$
$$
\tilde{R}_{i}=\text{diag}\left(\frac{1}{l_{i1}}R_{1},\cdots,\frac{1}{l_{iN}}R_{N}\right),
$$
$$
\tilde{P}_i = \Big(\sum_{j=1}^{N}l_{ij}P_{j}^{-1}\Big)^{-1},
$$
where the term $\frac{1}{l_{ij}}$ is set to be 0 if $l_{ij}=0$. Then the HCREs \eqref{HCRE} can be rewritten as
\begin{equation}\label{reHCRE}
\begin{aligned}
P_{i} &= A\Big(\tilde{P}_i^{-1}+\tilde{C}_{i}^{T}\tilde{R}_{i}^{-1}\tilde{C}_{i}\Big)^{-1}A^{T} + Q,\\
\tilde{P}_i &= \Big(\sum_{j=1}^{N}l_{ij}P_{j}^{-1}\Big)^{-1}.
\end{aligned}
\end{equation}
With the existence of the solution to HCREs \eqref{reHCRE}, for arbitrary one group of solution $\big\{P_i\big\}$, one can obtain
$$
\begin{aligned}
P_{i} &= A_{\tilde{P}_i}\tilde{P}_iA_{\tilde{P}_i}^T+Q+K_{\tilde{P}_i}\tilde{R}_{i}K_{\tilde{P}_i}^T,\\
A_{\tilde{P}_i} &= A-K_{\tilde{P}_i}\tilde{C}_{i},\\
K_{\tilde{P}_i} &= A\tilde{P}_i\tilde{C}_{i}^T\big(\tilde{C}_{i}\tilde{P}_i\tilde{C}_{i}^T+\tilde{R}_{i}\big)^{-1}.
\end{aligned}
$$
Moreover, there holds
$$
\begin{aligned}
\tilde{P}_i &= \tilde{P}_i\tilde{P}_i^{-1}\tilde{P}_i=\sum_{j=1}^{N}l_{ij}\tilde{P}_iP_j^{-1}\tilde{P}_i\\
&=\sum_{j=1}^{N}l_{ij}\tilde{P}_iP_j^{-1}P_jP_j^{-1}\tilde{P}_i\\
&=\sum_{j=1}^{N}l_{ij}\tilde{P}_iP_j^{-1}\big(A_{\tilde{P}_j}\tilde{P}_jA_{\tilde{P}_j}^T+Q+K_{\tilde{P}_j}\tilde{R}_{j}K_{\tilde{P}_j}^T\big)P_j^{-1}\tilde{P}_i.
\end{aligned}
$$
Let
$$
\begin{aligned}
\tilde{A}_{ij} &= \sqrt{l_{ij}}\tilde{P}_iP_j^{-1}A_{\tilde{P}_j},\\
Q_i&=\sum_{j=1}^{N}l_{ij}\tilde{P}_iP_j^{-1}QP_j^{-1}\tilde{P}_i,\\
\bar{R}_i&=\sum_{j=1}^{N}l_{ij}\tilde{P}_iP_j^{-1}K_{\tilde{P}_j}\tilde{R}_{j}K_{\tilde{P}_j}^TP_j^{-1}\tilde{P}_i.
\end{aligned}
$$
The expression of $\tilde{P}_i$ can be rewritten as
$$
\begin{aligned}
\tilde{P}_i&=\sum_{j=1}^{N}\tilde{A}_{ij}\tilde{P}_j \tilde{A}_{ij}^T+Q_i+\bar{R}_i\\
&=\sum_{j_1=1}^{N}\sum_{j_2=1}^{N}\tilde{A}_{ij_1}\tilde{A}_{j_1j_2}\tilde{P}_{j_2} \tilde{A}_{ij_1}^T\tilde{A}_{j_1j_2}^T+Q_i+\sum_{j=1}^{N}\tilde{A}_{ij}Q_j\tilde{A}_{ij}^T\\
&\quad+\bar{R}_i+\sum_{j=1}^{N}\tilde{A}_{ij}\bar{R}_j\tilde{A}_{ij}^T.
\end{aligned}
$$
In order to formulate the infinite series form of $\tilde{P}_i$ in a neat and compact form, let
$$
\begin{aligned}
\Phi_{i,j}^{(m)}\big(P\big)&=\sum_{j_1=1}^{N}\cdots\sum_{j_m=1}^{N}\tilde{A}_{ij_1}\tilde{A}_{j_1j_2}\cdots\tilde{A}_{j_mj}\times P\\
&\times\tilde{A}_{j_mj}^T\cdots\tilde{A}_{j_1j_2}^T\tilde{A}_{ij_1}^T,\qquad m>0,\\
\Phi_{i,j}^{(0)}\big(P\big)&=\tilde{A}_{ij}P\tilde{A}_{ij}^T.
\end{aligned}
$$
Then, the infinite series form of $\tilde{P}_i$ can be formulated as
\begin{equation}\label{seriesForm}
\begin{aligned}
\tilde{P}_i&=\sum_{j=1}^{N}\Phi_{i,j}^{(0)}\big(\tilde{P}_j\big)+Q_i+\bar{R}_i\\
&=\sum_{j=1}^{N}\Phi_{i,j}^{(1)}\big(\tilde{P}_j\big)+Q_i+\sum_{j=1}^{N}\Phi_{i,j}^{(0)}\big(Q_j\big)\\
&\quad+\bar{R}_i+\sum_{j=1}^{N}\Phi_{i,j}^{(0)}\big(\bar{R}_j\big)\\
&=\sum_{j=1}^{N}\Phi_{i,j}^{(m)}\big(\tilde{P}_j\big)+Q_i+\sum_{k=0}^{m-1}\sum_{j=1}^{N}\Phi_{i,j}^{(k)}\big(Q_j\big)\\
&\quad+\bar{R}_i+\sum_{k=0}^{m-1}\sum_{j=1}^{N}\Phi_{i,j}^{(k)}\big(\bar{R}_j\big).
\end{aligned}
\end{equation}
The following Lemma depicts the property of the operator $\Phi_{i,j}^{(m)}$.
\vspace{8pt}
\begin{lemma}\label{lmzero}
	For any positive definite matrix $P$, there holds 
	$$
	\lim\limits_{m\to\infty}\Phi_{i,j}^{(m)}\big(P\big)=\textbf{O}, \quad\forall i,j\in\mathcal{V}.
	$$
\end{lemma}
\vspace{6pt}
\begin{proof} Note that matrices $\Phi_{i,j}^{(m)}\big(\tilde{P}_j\big)$, $Q_i$, $\bar{R}_i$, $\Phi_{i,j}^{(k)}\big(Q_j\big)$ and $\Phi_{i,j}^{(k)}\big(\bar{R}_j\big)$ are positive definite. With the series form of $\tilde{P}$ in \eqref{seriesForm}, one can obtain
$$
\tilde{P}_i\ge Q_i+\sum_{k=0}^{m-1}\sum_{j=1}^{N}\Phi_{i,j}^{(k)}\big(Q_j\big).
$$
Due to the positive definiteness of $\Phi_{i,j}^{(k)}\big(Q_j\big)$, there is
$$
\lim\limits_{m\to\infty}\Phi_{i,j}^{(m)}\big(Q_j\big)=\textbf{O}, \quad\forall i,j\in\mathcal{V}.
$$
It is easy to verify that the operator $\Phi_{i,j}^{(m)}\big(\cdot\big)$ has the following two properties that 
$$
\Phi_{i,j}^{(m)}\big(\lambda P\big)=\lambda\Phi_{i,j}^{(m)}\big(P\big), \quad\forall i,j\in\mathcal{V},
$$
and for any positive definite matrices $P_1\ge P_2$, there is
$$
\Phi_{i,j}^{(m)}\big(P_1\big)\ge\Phi_{i,j}^{(m)}\big(P_2\big), \quad\forall i,j\in\mathcal{V}.
$$
For a fixed positive definite matrix $P$, one can choose a sufficient large number $\lambda$, such that $P\leq\lambda Q_j$, then one has
$$
\Phi_{i,j}^{(m)}\big(P\big)\leq\lambda\Phi_{i,j}^{(m)}\big(Q_j\big)\Rightarrow\lim\limits_{m\to\infty}\Phi_{i,j}^{(m)}\big(P\big)=\textbf{O},\quad\forall i,j\in\mathcal{V}.
$$ 
\end{proof}
\vspace{6pt}
The above Lemma shows that for any group of solution $\big\{P_i\big\}$ to the HCREs \eqref{HCRE}, the operator $\Phi_{i,j}^{(m)}(\cdot)$  converges to zero operator with the increase of $m$, which is essential for the following deduction. 

To fully excavate the property of the operator $\Phi_{i,j}^{(m)}\left(P\right)$, consider the following matrix 
$$
\begin{aligned}
\Big[\tilde{A}_{ij}\Big]^{(m)}\triangleq \Big[\tilde{A}_{ij_1}\times\tilde{A}_{j_1j_2}\cdots\times\tilde{A}_{j_mj},\cdots\Big],\\
j_1,\dots, j_m=1,\dots, N,
\end{aligned}
$$
which is a $n\times nN^m$ matrix and contains all the matrices $\tilde{A}_{ij_1}\tilde{A}_{j_1j_2}\cdots\tilde{A}_{j_mj}$ as row blocks, with $j_1,\dots, j_m$ chosen from $1$ to $N$. Hence, the operator $\Phi_{i,j}^{(m)}\big(P\big)$ can be rewritten as
$$
\Phi_{i,j}^{(m)}\big(P\big)=\Big[\tilde{A}_{ij}\Big]^{(m)}\big(I_{N^m}\otimes P\big)\Big(\Big[\tilde{A}_{ij}\Big]^{(m)}\Big)^T.
$$
Let $P=I$, with the the result in Lemma \ref{lmzero} that
$
\lim\limits_{m\to\infty}\Phi_{i,j}^{(m)}\big(I\big)=\textbf{O}, \;\;\forall i,j\in\mathcal{V},
$ one has 
$$
\lim\limits_{m\to\infty}\Big\|\Big[\tilde{A}_{ij}\Big]^{(m)}\Big\|_2=0, \quad\tau=1,2,\;\forall i,j\in\mathcal{V}.
$$
The asymptotic property of $\Big[\tilde{A}_{ij}\Big]^{(m)}$ with $m\to\infty$ will be repeatedly used in the following proof of the main theorems.

\vspace{6pt}
\begin{theorem}\label{thmunique}
	If Assumption \ref{assystem} and \ref{ascommunication} hold, the solution to \eqref{HCRE} is unique.
\end{theorem}
\vspace{6pt}
\begin{proof}
Based on Lemma \ref{lmexistence}, suppose there exist two different groups of solutions $\big\{P_{i}^{(1)}\big\}$ and $\big\{P_{i}^{(2)}\big\}$ to the HCREs \eqref{HCRE}. The corresponding notations are modified to be 
$$
\begin{aligned}
\tilde{P}_i^{(\tau)} = \Big(\sum_{j=1}^{N}l_{ij}\big(P_{j}^{(\tau)}\big)^{-1}\Big)^{-1},\quad\tau = 1,2.
\end{aligned}
$$
The corresponding modifications of $A_{\tilde{P}_i^{(\tau)}}$, $K_{\tilde{P}_i^{(\tau)}}$, $\tilde{A}^{(\tau)}_{ij}$, $Q_i^{(\tau)}$, $\bar{R}_i^{(\tau)}$ are made through the replacement of the matrix $P_i,\tilde{P}_i$ with $P_i^{(\tau)},\tilde{P}_i^{(\tau)}$.
The operator $\Phi_{i,j}^{(m)}(P)$ is also modified to $\Phi_{i,j}^{(m)(\tau)}(P)$, with the expression as 
$$
\Phi_{i,j}^{(m)(\tau)}\big(P\big)=\Big[\tilde{A}_{ij}^{(\tau)}\Big]^{(m)}\big(I_{N^m}\otimes P\big)\Big(\Big[\tilde{A}_{ij}^{(\tau)}\Big]^{(m)}\Big)^T.
$$
where
$$
\begin{aligned}
\Big[\tilde{A}_{ij}^{(\tau)}\Big]^{(m)}\triangleq \Big[\tilde{A}_{ij_1}^{(\tau)}\times\tilde{A}_{j_1j_2}^{(\tau)}\cdots\times\tilde{A}_{j_mj}^{(\tau)},\cdots\Big],\\
j_1,\dots, j_m=1,\dots N.
\end{aligned}
$$
A new operator is defined as
$$
\Psi_{i,j}^{(m)}\left(P\right)=\Big[\tilde{A}_{ij}^{(1)}\Big]^{(m)}\big(I_{N^m}\otimes P\big)\Big(\Big[\tilde{A}_{ij}^{(2)}\Big]^{(m)}\Big)^T,
$$
particularly with 
$$
\Psi_{i,j}^{(0)}\left(P\right)=\tilde{A}_{ij}^{(1)}P\big(\tilde{A}_{ij}^{(2)}\big)^T.
$$
Consider the difference between two groups of solutions, for any $i\in\mathcal{V}$, one has
$$
\begin{aligned}
&\tilde{P}_i^{(1)}-\tilde{P}_i^{(2)}=\tilde{P}_i^{(1)}\Big(\big(\tilde{P}_i^{(2)}\big)^{-1}-\big(\tilde{P}_i^{(1)}\big)^{-1}\Big)\tilde{P}_i^{(2)}\\
&=\sum_{j=1}^{N}l_{ij}\tilde{P}_i^{(1)}\big(P_j^{(1)}\big)^{-1}\big(P_j^{(1)}-P_j^{(2)}\big)\big(P_j^{(2)}\big)^{-1}\tilde{P}_i^{(2)}.
\end{aligned}
$$
With some calculations, one has
$$
\begin{aligned}
P_j^{(1)}-P_j^{(2)}&=A\Big(\big(\tilde{P}_i^{(1)}\big)^{-1}+\tilde{C}_{i}^{T}\tilde{R}_{i}^{-1}\tilde{C}_{i}\Big)^{-1}A^{T}\\
&\quad-A\Big(\big(\tilde{P}_i^{(2)}\big)^{-1}+\tilde{C}_{i}^{T}\tilde{R}_{i}^{-1}\tilde{C}_{i}\Big)^{-1}A^{T}\\
&=A_{\tilde{P}_j^{(1)}}\big(\tilde{P}_i^{(1)}-\tilde{P}_i^{(2)}\big)A_{\tilde{P}_j^{(2)}}^{T},
\end{aligned}
$$ 
where
$$
\begin{aligned}
A_{\tilde{P}_j^{(1)}}=A-A\tilde{P}_j^{(1)}\tilde{C}_{j}^T\big(\tilde{C}_{j}\tilde{P}_j^{(1)}\tilde{C}_{j}^T+\tilde{R}_{j}\big)^{-1}\tilde{C}_{j},\\
A_{\tilde{P}_j^{(2)}}=A-A\tilde{P}_j^{(2)}\tilde{C}_{j}^T\big(\tilde{C}_{j}\tilde{P}_j^{(2)}\tilde{C}_{j}^T+\tilde{R}_{j}\big)^{-1}\tilde{C}_{j}.
\end{aligned}
$$
The difference of $\tilde{P}_j^{(1)}$ and $\tilde{P}_j^{(2)}$ can be reformulated as
$$
\tilde{P}_j^{(1)}-\tilde{P}_j^{(2)}=\sum_{j=1}^{N}\Psi_{i,j}^{(0)}\big(\tilde{P}_j^{(1)}-\tilde{P}_j^{(2)}\big).
$$
With the property of $\Psi_{i,j}^{(m)}$, one has
$$
\tilde{P}_j^{(1)}-\tilde{P}_j^{(2)}=\sum_{j=1}^{N}\Psi_{i,j}^{(m)}\big(\tilde{P}_j^{(1)}-\tilde{P}_j^{(2)}\big),\quad m\in\mathbb{N}.
$$
Hence, to prove Theorem \ref{thmunique}, one only need to prove that for any matrix $P$, there is $\lim\limits_{m\to\infty}\Psi_{i,j}^{(m)}\big(P\big)=\textbf{O}$.

For any matrix $P$, there is
$$
\begin{aligned}
\Big\|\Psi_{i,j}^{(m)}\left(P\right)\Big\|_2&=\Big\|\Big[\tilde{A}_{ij}^{(1)}\Big]^{(m)}\big(I_{N^m}\otimes P\big)\Big(\Big[\tilde{A}_{ij}^{(2)}\Big]^{(m)}\Big)^T\Big\|_2\\
&\leq\Big\|\Big[\tilde{A}_{ij}^{(1)}\Big]^{(m)}\Big\|_2\Big\|\Big[\tilde{A}_{ij}^{(2)}\Big]^{(m)}\Big\|_2\big\|\big(I_{N^m}\otimes P\big)\big\|_2,
\end{aligned}
$$
where the inequality holds due to the fact that $\big\|\big(I_{N^m}\otimes P\big)\big\|_2=\big\|P\big\|_2$. 
Hence, with the discussion above Theorem \ref{thmunique}, for any matrix $P$, one has
$$
\lim\limits_{m\to\infty}\Big\|\Psi_{i,j}^{(m)}\left(P\right)\Big\|_2=0, \quad\forall i,j\in\mathcal{V}.
$$

Till now, the uniqueness of the solution $\big\{P_i\big\}$ is proved.
\end{proof}
\vspace{6pt}
\begin{remark}
	The above analysis demonstrates that when Assumptions \ref{assystem} and \ref{ascommunication} hold, a unique group of solutions exists for HCREs \eqref{HCRE}. This implies that the matrix iterative law of CI-based distributed filtering algorithm has a single fixed-point at the network level.  
	In contrast to the theory of traditional Riccati equations\cite{Andersonoptimal, kailath2000linear}, the local observability for each sensor needs not to be guaranteed for HCREs in this paper. As a result, the mathematical techniques used for traditional Riccati equations are not applicable for the analysis of HCREs. Additionally, due to the coupling between $P_i$, the series form of the solution $P_{i}$ is far more complex than traditional Riccati equations, which further complicates the proof of uniqueness.
	To overcome these challenges, we propose novel mathematical tools in this subsection, including the separation of $P_{i}$ and $\tilde{P}_i$, and the application of the operator $\Phi_{i,j}^{(m)}$. 
	However, the convergence of the iterative law to this fixed-point remains unsolved, and is the primary focus of the next subsection.
	
\end{remark}

\subsection{Iterative Law for Solving HCREs}\label{Convergence}
In this subsection, the convergence of the iterative law \eqref{HRiter} with respect to $k$ will be detailedly analyzed..

To simplify the proof, let $P_{i,k+1}$  represent the term $P_{i,k+1|k}$ in the iterative law of the CIDF Algorithm \ref{alg1}.

The iterative law \eqref{HRiter} can be reformulated as
$$
\begin{aligned}
P_{i,k+1} &= A\Big(\tilde{P}_{i,k}^{-1}+\tilde{C}_{i}^{T}\tilde{R}_{i}^{-1}\tilde{C}_{i}\Big)^{-1}A^{T} + Q,\\
\tilde{P}_{i,k+1} &= \Big(\sum_{j=1}^{N}l_{ij}P_{j,k+1}^{-1}\Big)^{-1}.
\end{aligned}
$$
With the similar technique proposed before, one can rewrite the iteration of $P_{i,k}$ as
$$
\begin{aligned}
P_{i,k+1} &= A_{\tilde{P}_{i,k}}\tilde{P}_{i,k}A_{\tilde{P}_{i,k}}^T+Q+K_{\tilde{P}_{i,k}}\tilde{R}_{i}K_{\tilde{P}_{i,k}}^T,\\
A_{\tilde{P}_{i,k}} &= A-K_{\tilde{P}_{i,k}}\tilde{C}_{i},\\
K_{\tilde{P}_{i,k}} &= A\tilde{P}_{i,k}\tilde{C}_{i}^T\big(\tilde{C}_{i}\tilde{P}_{i,k}\tilde{C}_{i}^T+\tilde{R}_{i}\big)^{-1}.
\end{aligned}
$$
Similarly, one can also rewrite the iteration of $\tilde{P}_{i,k+1}$ as
$$
\begin{aligned}
\tilde{P}_{i,k+1} &= \tilde{P}_{i,k+1}\tilde{P}_{i,k+1}^{-1}\tilde{P}_{i,k+1}\\
&=\sum_{j=1}^{N}l_{ij}\tilde{P}_{i,k+1}P_{j,k+1}^{-1}P_{j,k+1}P_{j,k+1}^{-1}\tilde{P}_{i,k+1}\\
&=\sum_{j=1}^{N}\tilde{A}_{ij,k+1}\tilde{P}_{j,k}\tilde{A}_{ij,k+1}^T+Q_{i,k+1}+\bar{R}_{i,k+1},
\end{aligned}
$$
where
$$
\begin{aligned}
\tilde{A}_{ij,k+1} &= \sqrt{l_{ij}}\tilde{P}_{i,k+1}P_{j,k+1}^{-1}A_{\tilde{P}_{j,k}},\\
Q_{i,k+1}&=\sum_{j=1}^{N}l_{ij}\tilde{P}_{i,k+1}P_{j,k+1}^{-1}QP_{j,k+1}^{-1}\tilde{P}_{i,k+1},\\
\bar{R}_{i,k+1}&=\sum_{j=1}^{N}l_{ij}\tilde{P}_{i,k+1}P_{j,k+1}^{-1}K_{\tilde{P}_{j,k}}\tilde{R}_{j}K_{\tilde{P}_{j,k}}^TP_{j,k+1}^{-1}\tilde{P}_{i,k+1}.
\end{aligned}
$$
To match the time-varying property of the iterative law, the original operator $\Phi_{i,j}^{(m)}$ is correspondingly modified as $\Phi_{i,j,k}^{(m)}$, where
$$
\begin{aligned}
\Phi_{i,j,k}^{(m)}\big(P\big)&=\Big[\tilde{A}_{ij,k}\Big]^{(m)}\Big(I_{N^{m}}\otimes P\Big)\times\Big(\Big[\tilde{A}_{ij,k}\Big]^{(m)}\Big)^T,\\
\Phi_{i,j,k}^{(0)}\big(P\big)&=\tilde{A}_{ij,k}P\tilde{A}_{ij,k}^T,
\end{aligned}
$$
with
$$
\begin{aligned}
\Big[\tilde{A}_{ij,k}\Big]^{(m)}\triangleq \Big[\tilde{A}_{ij_1,k}\times\tilde{A}_{j_1j_2,k-1}\cdots\times\tilde{A}_{j_{m}j,k-m},\cdots\Big],\\
j_1,\dots, j_{m}=1,\dots N.
\end{aligned}
$$
Then, one can also rewrite the iteration of $\tilde{P}_{i,k+1}$ as
$$
\begin{aligned}
\tilde{P}_{i,k+1}&=\sum_{j=1}^{N}\Phi_{i,j,k+1}^{(0)}\big(\tilde{P}_{j,k}\big)+Q_{i,k+1}+\bar{R}_{i,k+1}\\
&=\sum_{j=1}^{N}\Phi_{i,j,k+1}^{(m)}\big(\tilde{P}_{j,k-m}\big)+\sum_{h=0}^{m-1}\sum_{j=1}^{N}\Phi_{i,j,k+1}^{(h)}\big(Q_{j,k-h}\big)\\
&\quad+Q_{i,k+1}+\sum_{h=0}^{m-1}\sum_{j=1}^{N}\Phi_{i,j,k+1}^{(h)}\big(\bar{R}_{j,k-h}\big)+\bar{R}_{i,k+1}\\
&=\sum_{j=1}^{N}\Phi_{i,j,k+1}^{(k)}\big(\tilde{P}_{j,0}\big)+\sum_{h=0}^{k-1}\sum_{j=1}^{N}\Phi_{i,j,k+1}^{(h)}\big(Q_{j,k-h}\big)\\
&\quad+Q_{i,k+1}+\sum_{h=0}^{k-1}\sum_{j=1}^{N}\Phi_{i,j,k+1}^{(h)}\big(\bar{R}_{j,k-h}\big)+\bar{R}_{i,k+1},
\end{aligned}
$$
where $\tilde{P}_{i,0}^{-1}=\sum_{j=1}^{N}l_{ij}P_{j,0}^{-1}$ and $P_{j,0}$ is the initial value of iteration \eqref{HRiter}.
With Lemma \ref{GBlm}, the matrix $P_{i,k}$ is uniformly bounded for all $i\in\mathcal{V}$ and $k\ge\bar{k}$.
Hence, $\tilde{P}_{i,k}$ is also uniformly bounded for all $i\in\mathcal{V}$ and $k\ge\bar{k}$, and one has
$$
\tilde{P}_{i,k+1}\ge \sum_{j=1}^{N}\Phi_{i,j,k+1}^{(k)}\big(\tilde{P}_{j,0}\big).
$$

Similar to the deduction in the previous subsection, for any positive definite initial value $P_{i,0}$, there exists a positive number $M$ only related to $P_{i,0}$, such that
$$
\Big\|\Big[\tilde{A}_{ij,k}\Big]^{(k-1)}\Big\|_2\leq M,\quad\forall i,j\in\mathcal{V},k>0.
$$

With the proposed theoretical preparation, one can finally prove the convergence of iterative law \eqref{HRiter}.

\vspace{6pt}
\begin{theorem}\label{thmconverge}
	For any given matrices $A,C$, $Q$, $R$, $\mathcal{L}$ that satisfy Assumption \ref{assystem} and \ref{ascommunication}, the term $P_{i,k}$ of the iterative law \eqref{HRiter} converges to the unique solution $\big\{P_i\big\}$ of the HCREs \eqref{HCRE}, regardless of the initial value $P_{i,0}$, i.e., 
	$$\lim\limits_{k\to\infty}P_{i,k}=P_i,\quad\forall i\in\mathcal{V}.$$
\end{theorem}
\vspace{6pt}
\begin{proof}
One can first rewrite the iteration of the gap between $\tilde{P}_{i,k}$ and the unique solution $\tilde{P}_i$ as
$$
\begin{aligned}
&\tilde{P}_{i,k}-\tilde{P}_i=\tilde{P}_{i,k}\Big(\big(\tilde{P}_i\big)^{-1}-\big(\tilde{P}_{i,k}\big)^{-1}\Big)\tilde{P}_i\\
&=\sum_{j=1}^{N}l_{ij}\tilde{P}_{i,k}\big(P_{j,k}\big)^{-1}\big(P_{j,k}-P_j\big)\big(P_j\big)^{-1}\tilde{P}_i\\
&=\sum_{j=1}^{N}\tilde{A}_{ij,k}\big(\tilde{P}_{j,k-1}-\tilde{P}_j\big)\tilde{A}_{ij}^T\\
&=\sum_{j=1}^{N}\Big[\tilde{A}_{ij,k}\Big]^{(k-1)}\Big(I_{N^{k-1}}\otimes \big(\tilde{P}_{j,0}-\tilde{P}_j\big)\Big)\Big(\Big[\tilde{A}_{ij}\Big]^{(k-1)}\Big)^T.
\end{aligned}
$$
where the definition of $\Big[\tilde{A}_{ij}\Big]^{(k-1)}$ is the same as that of Subsection \ref{Uniqueness}. Hence, one can obtain the norm of the gap as
$$
\begin{aligned}
\big\|\tilde{P}_{i,k}-\tilde{P}_i\big\|_2&\leq\sum_{j=1}^{N} \Big\|\Big[\tilde{A}_{ij,k}\Big]^{(k-1)}\Big\|_2\big\|\tilde{P}_{i,0}-\tilde{P}_i\big\|_2\\
&\quad\times \Big\|\Big[\tilde{A}_{ij}\Big]^{(k-1)}\Big\|_2.
\end{aligned}
$$
For any initial value $\tilde{P}_{i,0}$, due to the uniform boundedness of $\Big\|\Big[\tilde{A}_{ij,k}\Big]^{(k-1)}\Big\|_2$ and the fact that 
$$
\lim\limits_{k\to\infty}\Big\|\Big[\tilde{A}_{ij}\Big]^{(k-1)}\Big\|_2=0, \quad\forall i,j\in\mathcal{V}.
$$
One can finally obtain that 
$$
\lim\limits_{k\to\infty}\big\|\tilde{P}_{i,k}-\tilde{P}_i\big\|_2=0,
$$
for any initial value $\tilde{P}_{i,0}$. The convergence of the iteration \eqref{HRiter} is finally proved.
\end{proof}
\vspace{6pt}
\begin{remark}
	Theorem \ref{thmconverge} demonstrates that the {\it a priori} covariance matrix term of CI-based distributed filtering $P_{i,k+1|k}$, converges to the unique steady-state performance $P_{i}$, as $k$ tends to infinity, regardless of the initial value. Compared to previous literature \cite{battistelli2014kullback, battistelli20156897960, he2018consistent, he2020distributed, wang2017convergence} that only proves the boundedness of $P_{i,k+1|k}$, Theorem \ref{thmconverge} further confirms the convergence of $P_{i,k+1|k}$. This result is significant as it establishes a concise relationship between filtering performance and the filtering parameter $\mathcal{L}$ with HCREs \eqref{HCRE}, reducing the conservativeness of the performance evaluation. The steady-state performance of $P_{i,k+1|k}$ also motivates the formulation of the steady-state performance of the real estimation error covariance matrix, which will be proposed in the next section.
\end{remark}
\vspace{6pt}

The numerical example presented below demonstrates that the solution of HCREs \eqref{HCRE} is significantly smaller than the upper bound proposed in the literature of CI-based distributed filtering, such as \cite{battistelli2014kullback, battistelli20156897960}. This result confirms that the performance evaluation of CI-based distributed filtering is much less conservative than traditional techniques. 

Consider an one-dimension system with $A=1,\;C_1=1,\;C_2=C_3=0,\;Q=1,\;R=1$ and
$$
\mathcal{L}=\begin{bmatrix}
\frac{1}{2}&\frac{1}{2}&0\\
\frac{1}{3}&\frac{1}{3}&\frac{1}{3}\\
0&\frac{1}{2}&\frac{1}{2}\\
\end{bmatrix}.
$$
Through performing the iterative law \eqref{HRiter} for sufficient many times, one can obtain the solution of HCREs \eqref{HCRE} as $P_1=2.0492,\;P_2=2.3909,\;P_3=3.9901$. With the method proposed in \cite{battistelli2014kullback}, one has
$$
\begin{aligned}
P_{3,k+1|k}&=AP_{3,k}A^T+Q\\
&= A\big(0.5P_{2,k|k-1}^{-1}+0.5P_{3,k|k-1}^{-1}\big)^{-1}A^T+Q\\
&\leq A\big(0.5\beta A^{-T}P_{2,k}^{-1}A^{-1}\big)^{-1}A^T+Q\\
&\leq A\Big(0.5\beta A^{-T}\big(\frac{1}{3}C_1^TR^{-1}C_1\big)A^{-1}\Big)^{-1}A^T+Q,
\end{aligned}
$$
where $\beta< 1$. In this case, one has  $P_{3,k+1|k}\ge 7$, which is larger than the exact solution $P_{3}=3.9901$. This indicates that the classical performance evaluation technique is conservative and the estimated values are much higher than the actual values. On the other hand, HCREs provides a solution that is much smaller than the upper bound proposed in the literature for CI-based distributed filtering, which means that it is less conservative and provides more accurate evaluation.
\section{The Application of HCREs in CI-based Distributed Filtering}\label{secdiscussion}
In this section, some new perspectives of the CI-based distributed filtering algorithm are proposed on the basis of the HCREs theory obtained in the previous section.
\subsection{Steady-State Performance of the Real Covariance Matrix}\label{seccovariance}
It is mentioned that the parameter matrix $P_{i,k+1|k}$ converges to the steady-state form with the increase of $k$ regardless of the initial value. In this subsection, the corresponding steady-state performance of the real estimation error covariance matrix will be formulated.
Consider the LTI system proposed in Subsection \ref{subModel} and the CI-based distributed filtering algorithm proposed in Algorithm \ref{alg1}.
The estimation error is defined as
$$
\begin{aligned}
e_{i,k|k-1}&=x_{k} - \hat{x}_{i,k|k-1},\\
e_{i,k|k}&=x_{k} - \hat{x}_{i,k|k},\\
e_{i,k}&=x_k-\hat{x}_{i,k},
\end{aligned}
$$
and iteration of the estimation error can be reformulated as 
$$
\begin{aligned}
e_{i,k+1|k}=&\sum_{j=1}^Nl_{ij}AP_{i,k}P_{j,k|k-1}^{-1}e_{j,k|k-1}+\omega_k\\
&+\sum_{j=1}^Nl_{ij}AP_{i,k}C_j^TR_j^{-1}v_{j,k}.
\end{aligned}
$$
Denote the following notations as
$$
\begin{aligned}
\mathcal{A}_k&\triangleq\begin{bmatrix}
l_{11}AP_{1,k}P_{1,k|k-1}^{-1}&\cdots&l_{1N}AP_{1,k}P_{N,k|k-1}^{-1}\\
\vdots&\ddots&\vdots\\
l_{N1}AP_{N,k}P_{1,k|k-1}^{-1}&\cdots&l_{NN}AP_{N,k}P_{N,k|k-1}^{-1}
\end{bmatrix},\\
\Gamma_k&\triangleq\begin{bmatrix}
l_{11}AP_{1,k}C_1R_1^{-1}&\cdots&l_{1N}AP_{1,k}C_NR_N^{-1}\\
\vdots&\ddots&\vdots\\
l_{N1}AP_{N,k}C_1R_1^{-1}&\cdots&l_{NN}AP_{N,k}C_NR_N^{-1}
\end{bmatrix},
\end{aligned}
$$
and
$$
\begin{aligned}
e_{k+1|k}&\triangleq\begin{bmatrix}
e_{1,k+1|k}^T&\cdots&e_{N,k+1|k}^T
\end{bmatrix}^T,\\
v_{k}&\triangleq\begin{bmatrix}
v_{1,k}^T&\cdots&v_{N,k}^T
\end{bmatrix}^T.
\end{aligned}
$$
Then, the iterative law of $e_{k+1|k}$ can be reformulated as a compact form
$$
e_{k+1|k}=\mathcal{A}_ke_{k|k-1}+\Gamma_kv_{k}+\textbf{1}_N\otimes\omega_k.
$$
As the matrix $P_{i,k}$ and $P_{j,k|k-1}$ converge to the steady-state form with $k$ tending to infinity, i.e.,
$$
\lim\limits_{k\to\infty}P_{i,k}=\bar{P}_{i},\quad\lim\limits_{k\to\infty}P_{i,k|k-1}=P_{i}.
$$
where
$$
\bar{P}_i=\Big(\tilde{C}_i^T\tilde{R}_i^{-1}\tilde{C}_i+\sum_{j=1}^{N}l_{ij}P_j^{-1}\Big)^{-1},
$$
and the steady-state matrices satisfy $A\bar{P}_{i}A^T=P_{i}-Q\leq \beta P_{i},\;\; \forall i\in\mathcal{V}$, where $0<\beta<1$ due to the positive definiteness of $Q$.

With Theorem \ref{thmconverge}, the matrices $\mathcal{A}_k$ and $\Gamma_k$ will also converge to the steady-state form with the increase of $k$, i.e.,
$$
\begin{aligned}
&\lim\limits_{k\to\infty}\mathcal{A}_k=\mathcal{A}\triangleq\begin{bmatrix}
l_{11}A\bar{P}_1P_{1}^{-1}&\cdots&l_{1N}A\bar{P}_1P_{N}^{-1}\\
\vdots&\ddots&\vdots\\
l_{N1}A\bar{P}_NP_{1}^{-1}&\cdots&l_{NN}A\bar{P}_NP_{N}^{-1}
\end{bmatrix},\\
&\lim\limits_{k\to\infty}\Gamma_k=\Gamma\triangleq\begin{bmatrix}
l_{11}A\bar{P}_{1}C_1R_1^{-1}&\cdots&l_{1N}A\bar{P}_{1}C_NR_N^{-1}\\
\vdots&\ddots&\vdots\\
l_{N1}A\bar{P}_{N}C_1R_1^{-1}&\cdots&l_{NN}A\bar{P}_{N}C_NR_N^{-1}
\end{bmatrix}.
\end{aligned}
$$ 
Consider the Perron-Frobenius left eigenvector $q$ of the stochastic matrix $\mathcal{L}$, which satisfies $q^T\mathcal{L}=q^T$, and the matrix 
$$
\mathcal{Q}=\text{diag}\left(q_1P_1^{-1},\cdots,q_NP_N^{-1}\right),
$$ 
which is a block diagonal matrix.
Note that
$$
\begin{bmatrix}
q_1A^TP_1^{-1}A& & \\
&\ddots& \\
& & q_NA^TP_N^{-1}A
\end{bmatrix}\leq
\beta\bar{\mathcal{Q}},
$$
where 
$$
\bar{\mathcal{Q}}=\begin{bmatrix}
q_1\bar{P}_1^{-1}& & \\
&\ddots& \\
& & q_N\bar{P}_N^{-1}
\end{bmatrix}.
$$
Then, one has
$$
\begin{aligned}
\mathcal{A}^T\mathcal{Q}\mathcal{A}=&\begin{bmatrix}
l_{11}A\bar{P}_1P_{1}^{-1}&\cdots&l_{1N}A\bar{P}_1P_{N}^{-1}\\
\vdots&\ddots&\vdots\\
l_{N1}A\bar{P}_NP_{1}^{-1}&\cdots&l_{NN}A\bar{P}_NP_{N}^{-1}
\end{bmatrix}^T
\times\mathcal{Q}\star\\
\leq&\beta\begin{bmatrix}
l_{11}\bar{P}_1P_{1}^{-1}&\cdots&l_{1N}\bar{P}_1P_{N}^{-1}\\
\vdots&\ddots&\vdots\\
l_{N1}\bar{P}_NP_{1}^{-1}&\cdots&l_{NN}\bar{P}_NP_{N}^{-1}
\end{bmatrix}^T
\times\bar{\mathcal{Q}}\star\\
=&\beta\sum_{i=1}^{N}\begin{bmatrix}
l_{i1}P_1^{-1}\\ \vdots\\
l_{iN}P_N^{-1}
\end{bmatrix}q_i\bar{P}_i
\begin{bmatrix}
l_{i1}P_1^{-1}& \cdots&
l_{iN}P_N^{-1}
\end{bmatrix},
\end{aligned}
$$
where the term $\star$ denotes the transpose of corresponding matrices.
With the fact that $\bar{P}_i\leq\tilde{P}_i\triangleq\big(\sum_{j=1}^Nl_{ij}P_j^{-1}\big)^{-1}$ and the Lemma that (equivalent to Lemma 2 in \cite{battistelli2014kullback})
$$
\begin{bmatrix}
P_1\\ \vdots\\
P_N
\end{bmatrix}\big(\sum_{j=1}^NP_j\big)^{-1}
\begin{bmatrix}
P_1& \cdots&
P_N
\end{bmatrix}\leq\begin{bmatrix}
P_1& & \\
&\ddots& \\
& &P_N
\end{bmatrix},
$$
one can obtain that
$$
\mathcal{A}^T\mathcal{Q}\mathcal{A}\leq\beta\sum_{i=1}^{N}q_i\begin{bmatrix}
l_{i1}P_1^{-1}& & \\
&\ddots& \\
& & l_{iN}P_N^{-1}
\end{bmatrix}=\beta\mathcal{Q}.
$$
Due to the positive definiteness of $\mathcal{Q}$, one can obtain that the matrix $\mathcal{A}$ is Schur stable. With the Theorem 1 in \cite{cattivelli2010diffusion}, the estimation error covariance matrix will converge to the solotion of the Lyapunov equation, i.e.,
$$
\begin{aligned}
\lim\limits_{k\to\infty}\mathbb{E}\left\{e_{k+1|k}e_{k+1|k}^T\right\}=\mathcal{P},
\end{aligned}
$$
where
\begin{equation}\label{DLE}
\mathcal{P}=\mathcal{A}\mathcal{P}\mathcal{A}^T+\Gamma R\Gamma^T+\textbf{1}_N\textbf{1}_N^T\otimes Q.
\end{equation}

\begin{remark}
	In the matrix $\mathcal{A}_k$, the term $l_{ij}AP_{i,k}P_{j,k|k-1}^{-1}$ is a crucial element to consider. It is important to note that if $P_{j,k|k-1}$ is large, this term will be relatively small. This means that the information weight of sensor $j$ will be reduced, which is an essential factor in ensuring the stability of the CI-based distributed filter even without local observability.
	Compared with the literature \cite{battistelli2014kullback, battistelli20156897960, wang2017convergence}, 
	the analysis in this subsection provides a new formulation for the steady-state performance of the estimation error covariance matrix of the CIDF algorithm. Specifically, we have derived the explicit form of a discrete-time Lyapunov equation 
	\eqref{DLE}, with the parameter matrices of the DLE obtained through solving the HCREs \eqref{HCRE}. 
    This new result establishes a quantitative relationship between the filtering performance and the weighted parameter matrix $\mathcal{L}$, providing essential performance metric for optimizing the parameters of CIDF algorithms.	
\end{remark}
\subsection{Unification of CI-Based Distributed Filtering with HCREs}
This subsection will examine the structural similarities of several widely recognized CIDF algorithms proposed in \cite{battistelli2014kullback, kamal2013information, battistelli20156897960} using the HCREs framework. It will be demonstrated that the steady-state performances of all distributed algorithms based on CI can be unified as a solution to a discrete-time Lyapunov equation. By solving the corresponding HCREs \eqref{HCRE}, the parameter matrices of the DLE can be obtained.

Consider the following three matrix iterative law proposed in \cite{battistelli2014kullback, kamal2013information, battistelli20156897960}, respectively.
 The first one is the basic CI-based distributed matrix iterative law proposed in \cite{battistelli2014kullback}, i.e.,
$$
\begin{aligned}
&P_{i,k+1|k} = AP_{i,k}A^{T} + Q,\\
&P_{i,k} = \Big(\sum_{j=1}^{N}l_{ij}P_{j,k|k-1}^{-1}+\l_{ij}C_j^TR_j^{-1}C_j\Big)^{-1}.
\end{aligned}
$$
and the estimation iterative law
$$
\begin{aligned}
\hat{x}_{i,k+1|k}&=AP_{i,k}\Big(\sum_{j=1}^{N}l_{ij}P_{j,k|k-1}^{-1}\hat{x}_{j,k|k-1}+l_{ij}C_jR_j^{-1}y_{j,k}\Big).\\
\end{aligned}
$$

The second one is the matrix iterative law of information weighted consensus filters proposed in \cite{kamal2013information} (equation (33), (35) and (39) of \cite{kamal2013information}), i.e.,
	$$
	\begin{aligned}
	&P_{i,k+1|k} = AP_{i,k}A^{T} + Q,\\
	&P_{i,k} = \Big(\sum_{j=1}^{N}l_{ij}^{(L)}P_{j,k|k-1}^{-1}+\l_{ij}^{(L)}NC_j^TR_j^{-1}C_j\Big)^{-1},
	\end{aligned}
	$$
where $l_{ij}^{(L)}$ is the $\big(i,j \big)$-th element of $\mathcal{L}^{L}$, $\mathcal{L}$ takes a special form as $\mathcal{L}=I-\epsilon\bar{\mathcal{L}}$ and $N$ is the number of the nodes. 
The estimation iterative law is proposed as
\begin{small}
	$$
	\begin{aligned}
	&\hat{x}_{i,k+1|k}=AP_{i,k}\Big(\sum_{j=1}^{N}l_{ij}^{(L)}P_{j,k|k-1}^{-1}\hat{x}_{j,k|k-1}+l_{ij}^{(L)}NC_jR_j^{-1}y_{j,k}\Big).
	\end{aligned}
	$$
\end{small}

The third one is the matrix iterative law of the hybrid CMCI algorithm proposed in \cite{battistelli20156897960}, i.e.,

	$$
	\begin{aligned}
	&P_{i,k+1|k} = AP_{i,k}A^{T} + Q,\\
	&P_{i,k} = \Big(\sum_{j=1}^{N}l_{ij}^{(L)}P_{j,k|k-1}^{-1}+\l_{ij}^{(L)}\omega_{j,k}C_j^TR_j^{-1}C_j\Big)^{-1},
	\end{aligned}
	$$
where $l_{ij}^{(L)}$ is also the $\big(i,j\big)$-th element of $\mathcal{L}^{L}$ and $\omega_{j,k}$ is the local weight paramter determined by each sensor node $j$.
The estimation iterative law is proposed as
\begin{footnotesize}
	$$
	\hat{x}_{i,k+1|k}=AP_{i,k}\Big(\sum_{j=1}^{N}l_{ij}^{(L)}P_{j,k|k-1}^{-1}\hat{x}_{j,k|k-1}+l_{ij}^{(L)}\omega_{j,k}C_jR_j^{-1}y_{j,k}\Big).
	$$
\end{footnotesize}

All of the above mentioned matrix iterative laws of $P_{i,k+1|k}$ can be simplyfied as the following HCREs matrix iterative law with different paramter matrices
\begin{equation}\label{mHCRE}
P_{i,k+1|k} = A\Big(\sum_{j=1}^{N}l_{ij}P_{j,k|k-1}^{-1}+\nu_{ij}C_j^TR_j^{-1}C_j\Big)^{-1}A^{T} + Q,
\end{equation}
where $\mathcal{L}\triangleq \big\{l_{ij}\big\}$, $\nu\triangleq \big\{\nu_{ij}\big\}$ are parameter matrices corresponding with the communication graph $\mathcal{G}$. 
In addition, the estimation iterative law can also be unified as
$$
\hat{x}_{i,k+1|k}=AP_{i,k}\Big(\sum_{j=1}^{N}l_{ij}P_{j,k|k-1}^{-1}\hat{x}_{j,k|k-1}+\nu_{ij}C_jR_j^{-1}y_{j,k}\Big).
$$
Both of $\mathcal{L}$ and $\nu$ are irreducible and $\mathcal{L}$ is row-stochastic. Based on the above analysis, the property of such an iterative law is mainly discussed in this subsection.

Let $\mathcal{L}^{(m)}\triangleq \big\{l_{ij}^{(m)}\big\}$, $\varpi^{(m)}\triangleq \mathcal{L}^{(m-1)}\nu\triangleq\big\{\varpi_{ij}^{(m)}\big\}$, where $\mathcal{L}^{(0)}=I$ and $\varpi^{(1)}=\nu$.
With Lemma 1 in \cite{battistelli2014kullback}, one can obtain that 
$$
\begin{aligned}
P_{i,k+1}^{-1}&\ge \beta A^{-T}\Big(\sum_{j=1}^{N}l_{ij}P_{j,k}^{-1}+\nu_{ij}C_j^TR_j^{-1}C_j\big)A^{-1}\\
&\ge\beta^{m}\sum_{j=1}^{N}l_{ij}^{(m)}\big(A^{-1}\big)^mP_{j,k}^{-1}\big(A^{-T}\big)^m\\
&\quad+\sum_{k=1}^{m}\beta^{k}\sum_{j=1}^{N}\varpi_{ij}^{(k)}\big(A^{-1}\big)^{k}C_j^TR_j^{-1}C_j\big(A^{-T}\big)^{k},
\end{aligned}
$$
where $\beta<1$. Note that both of $\mathcal{L}$ and $\nu$ are irreducible, one has that for any $k\ge N$, there is $\varpi_{ij}^{(k)}>0$, $l_{ij}^{(k)}>0$ for any $i,j\in\mathcal{V}$. Together with the invertibility of matrix $A$ and the observability of $\big(A,C\big)$, one can obtain that the term $P_{i,k}$ of matrix iterative law \eqref{mHCRE} is uniformly bounded, i.e., there exists a matrix $P$ and a sufficient large number $\bar{k}$, such that $P_{i,k}\leq P,\;\forall i\in\mathcal{V}, k\ge\bar{k}$. 

With the uniform boundedness of the term $P_{i,k+1|k}$ in \eqref{mHCRE}, similar to the derivation procedure of Section \ref{Uniqueness} and \ref{Convergence}, one 
has the follwing lemma:
\vspace{6pt}
\begin{lemma}
	For any given matrices $A,C$, $Q$, $R$, $\mathcal{L}$ that satisfy Assumption \ref{assystem} and \ref{ascommunication}, if $\nu$ is primitive, the modified HCREs,
	\begin{equation}\label{steady}
	P_{i} = A\Big(\sum_{j=1}^{N}l_{ij}P_{j}^{-1}+\nu_{ij}C_j^TR_j^{-1}C_j\Big)^{-1}A^{T} + Q,
	\end{equation}
	has a unique group of solution $\big\{P_i\big\}$. Moreover, the term $P_{i,k}$ of iterative law \eqref{mHCRE} converge to the unique solution $P_i$ with $k\to\infty$ for any $i\in\mathcal{V}$.
\end{lemma}
\vspace{6pt}

In light of the analysis presented above, it can be concluded that all of the matrix iterative laws proposed in \cite{battistelli2014kullback, kamal2013information, battistelli20156897960} can be viewed as the same iterative law \eqref{mHCRE} with different parameter matrices $\nu$. This implies that the differences between the three information fusion mechanisms are ultimately reflected in the values of the parameter matrix $\nu$. Furthermore, the steady-state performances $P_i$ of all three iterative laws can be simplified as the solution to the modified HCREs \eqref{steady}.

As for the real estimation error covariance matrix, with the unified estimation iterative law, the iterative law of estimation error can be correspondingly modified as
$$
\begin{aligned}
e_{i,k+1|k}=&\sum_{j=1}^Nl_{ij}AP_{i,k}P_{j,k|k-1}^{-1}e_{j,k|k-1}+\omega_k\\
&+\sum_{j=1}^N\nu_{ij}AP_{i,k}C_j^TR_j^{-1}v_{j,k}.
\end{aligned}
$$
Hence, the steady state performance of all kinds of CI-based distributed filtering algorithm can also be simplyfied as the solution $\mathcal{P}$ to the following discrete-time Lyapunov equation 
\begin{equation}\label{mDLE}
\mathcal{P}=\mathcal{A}\mathcal{P}\mathcal{A}^T+\Gamma R\Gamma^T+\textbf{1}_N\textbf{1}_N^T\otimes Q,
\end{equation}
where $\Gamma$ is modified to be 
$$
\Gamma\triangleq\begin{bmatrix}
\nu_{11}A\bar{P}_{1}C_1R_1^{-1}&\cdots&\nu_{1N}A\bar{P}_{1}C_NR_N^{-1}\\
\vdots&\ddots&\vdots\\
\nu_{N1}A\bar{P}_{N}C_1R_1^{-1}&\cdots&\nu_{NN}A\bar{P}_{N}C_NR_N^{-1}
\end{bmatrix}
$$

\subsection{Asymptotic Analysis of Filtering Performance with Paramters $L$ and $\mathcal{L}$}
This subsection will delve into the asymptotic properties of HCREs. Specifically, the matrix performances of various CI-based filtering algorithms as $L\to\infty$. It is assumed that the parameter matrices $\mathcal{L}$ and $\nu$ are row stochastic to obtain the necessary asymptotic results for sufficiently large fusion steps $L$.

For primitive and row stochastic matrices $\mathcal{L}$ and $\nu$, there exist vectors $\mu_1,\mu_2,\mu_3,\mu_4$, such that 
$$
\begin{aligned}
\mathcal{L}\mu_1=\mu_1,\quad \mu_2^T\mathcal{L}=\mu_2^T,\quad\nu \mu_3=\mu_3,\quad \mu_4^T\nu=\mu_4^T.
\end{aligned}
$$
where $\mu_2^T\mu_1=\mu_4^T\mu_3=1$.
From Theorem 8.5.1 in \cite{Horn1985}, one has
$$
\lim\limits_{k\to\infty}\mathcal{L}^k=\mu_1\mu_2^T,\qquad\lim\limits_{k\to\infty}\nu^k=\mu_3\mu_4^T.
$$
Meanwhile, since $\mathcal{L}$ and $\nu$ are row stochastic, one has $\mu_1=\mu_3=\textbf{1}_N$, and
$$\lim\limits_{k\to\infty}l_{ij}^{(k)}=\mu_{2,j},\qquad \lim\limits_{k\to\infty}\nu_{ij}^{(k)}=\mu_{4,j},$$
where $\mu_{2,j},\mu_{4,j}$ is the $j$-th element of vectors $\mu_2,\mu_4$, respectively.

Through performing the information fusion step for $L$ times, one can rewrite the modified HCREs \eqref{mHCRE} as
$$
P_{i,k+1|k} = A\Big(\sum_{j=1}^{N}l_{ij}^{(L)}P_{j,k|k-1}^{-1}+\nu_{ij}^{(L)}C_j^TR_j^{-1}C_j\Big)^{-1}A^{T} + Q,
$$
and the aysmptotic form of iteration law with $L\to\infty$ as
$$
P_{i,k+1|k} = A\Big(\sum_{j=1}^{N}\mu_{2,j}P_{j,k|k-1}^{-1}+\mu_{4,j}C_j^TR_j^{-1}C_j\Big)^{-1}A^{T} + Q.
$$
Note that with $L\to\infty$, the iteration law of each sensor $i$ will tend to be the same with each other due to the fact that $\mathcal{L}$ and $\nu$ are row stochastic. Therefore, even if the initial value $P_{i,0}$ is different with respect to $i$, there is $P_{i,k+1|k}=P_{j,k+1|k},\forall i,j\in\mathcal{V},k\ge 1$. With the property that $\mu_2^T\mu_1=\mu_2^T\textbf{1}_N=1$, one can further rewrite the above iteration law as

\begin{equation}\label{Asymptotic}
P_{i,k+1|k} = A\Big(P_{i,k|k-1}^{-1}+\sum_{j=1}^{N}\mu_{4,j}C_j^TR_j^{-1}C_j\Big)^{-1}A^{T} + Q.
\end{equation}
With the obatined aysmptotic iterative law \eqref{Asymptotic}, one can further compare the asymptotic performance of the consensus-based distributed filtering algorithm proposed in \cite{battistelli2014kullback, kamal2013information, battistelli20156897960}, respectively.

For the iterative law proposed in \cite{battistelli2014kullback}, for the special case that the matrix $\nu$ is doubly stochastic, $\mu_4=\frac{1}{N}\textbf{1}_N$. Hence, with the fusion step $L$ tending to infinity, the asymptotic form of the iterative law can be rewritten as
$$
P_{i,k+1|k} = A\Big(P_{i,k|k-1}^{-1}+\frac{1}{N}\sum_{j=1}^{N}C_j^TR_j^{-1}C_j\Big)^{-1}A^{T} + Q.
$$ 

For the iterative law proposed in \cite{kamal2013information}, with the parameter $N$ to compensate for the underconfidence of the information matrix, one can finally rewrite the asymptotic iterative law as 
$$
P_{i,k+1|k} = A\Big(P_{i,k|k-1}^{-1}+\sum_{j=1}^{N}C_j^TR_j^{-1}C_j\Big)^{-1}A^{T} + Q.
$$
Hence, with sufficiently large fusion step $L$, the performance of the IKF proposed in \cite{kamal2013information} converges to the centralized optimal performance.

For the iterative law proposed in \cite{battistelli20156897960}, for the special case that the matrix $\nu$ is doubly stochastic, one can finally rewrite the asymptotic iterative law as 
$$
P_{i,k+1|k} = A\Big(P_{i,k|k-1}^{-1}+\frac{1}{N}\sum_{j=1}^{N}\omega_{j,k}C_j^TR_j^{-1}C_j\Big)^{-1}A^{T} + Q,
$$
where the parameter $\omega_{j,k}$ can be tuned by each sensor node. 

Based on the analysis presented above, it is evident that the performance gap between the CI-based filtering algorithm proposed in \cite{battistelli2014kullback} and the centralized optimal case increases significantly with large sensor numbers $N$ and fusion steps $L$. This phenomenon has also been discussed in \cite{battistelli2014kullback}, where it was emphasized that the information fusion operation in the CI-based filtering algorithm requires a cautious strategy to ensure robustness against data incest (i.e., the repeated usage of the same observation). The compensation strategy proposed in \cite{kamal2013information} can ensure asymptotic optimality of filtering performance, but for finite fusion steps $L$, the algorithm may suffer from inconsistency induced by overconfidence in the observation information, where $N\nu_{ij}^{(L)}$ is much larger than 1. Therefore, future research should focus on developing effective parameter tuning techniques to maintain a proper balance between the information fusion weights and estimation performance.

\section{Simulation}\label{secsimulation}
    This section includes two numerical experiments aimed at validating the theory proposed in this paper. The first experiment verifies the HCREs theory presented in Section \ref{sec3}, which contains proving the uniqueness of the solution to HCREs and demonstrating the convergence of the iterative law. The second experiment validates the theory proposed in Section \ref{secdiscussion}, which contains the explicit form of the steady-state performance of the error covariance matrix of the CIDF algorithm and a performance comparison between different filtering algorithms.
    
	In the first numerical experiment, in order to fully validate the theory of HCREs, the matrices $A,C,Q,R$ are all randomly generated, with the expression as
	$$
	\begin{aligned}
    A=\begin{bmatrix}
	0.3836&	0.2558&	0.2525&	0.1766&	0.4524&	0.3534\\
	0.1978&	0.2351&	0.4546&	0.5642&	0.1793&	0.4899\\
	0.3322&	0.4508&	0.4779&	0.4064&	0.5716&	0.4073\\
	0.5927&	0.4560&	0.5109&	0.6161&	0.2135&	0.1504\\
	0.6139&	0.4898&	0.3574&	0.3858&	0.6741&	0.6985\\
	0.5016&	0.0795&	0.0191&	0.5526&	0.0543&	0.4081
	\end{bmatrix}.
	\end{aligned}
	$$
	The maximum eigenvalue of $A$ is $2.31$, which indicates that the matrix $A$ is not schur stable.
	
	There are three kinds of the observation matrices:
	$$
	\begin{aligned}
	&C^{(1)}=\left[0.3711,0.4438,0.2733,0.3920,0.3768,0.1424\right],\\
	&C^{(2)}=\left[0.7154,0.3439,0.4017,0.9339,0.1471,0.2543\right],\\
	&C^{(3)}=\left[0,0,0,0,0,0\right].
	\end{aligned}
	$$
	
\begin{figure}
		\centering
		\includegraphics[width=0.5\textwidth]{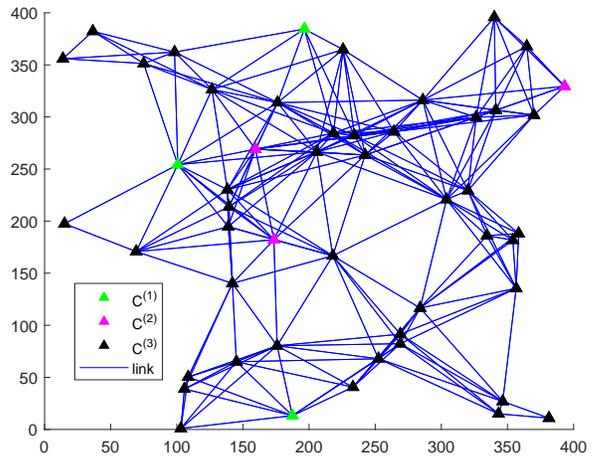}
		\caption{
			Illustration figure for communication topology corresponding to the matrix $\mathcal{L}$ in HCREs \eqref{HCRE}.}
		\label{communication}	
	\end{figure}

	The value of $\mathcal{L}$ is obtained with a randomly generated communication  network. The whole network corresponding to the weighting matrix $\mathcal{L}$ in HCREs \eqref{HCRE} consists of 50 nodes, including 3 nodes of kind $C^{(1)}$, 3 nodes of kind $C^{(2)}$, 44 nodes of kind $C^{(3)}$. {The locations of the nodes are randomly set in a $500\times500$ {region} and each node is with a communication radius of $110$.} Hence, the communication topology of the network and the structure of weighting matrix $\mathcal{L}$ are randomly generated in the numerical experiments, {as presented in Fig.~\ref{communication}.}

    The covariance matrix $Q$ and $R_i$ takes the form as
    $$
    \begin{aligned}
    Q&=\begin{bmatrix}
    1.79&   -0.69&    0.48&   -0.39&   -0.26&   -0.25\\
    -0.69&    1.45&   -0.07&    0.01&    0.56&    0.05\\
    0.48&   -0.07&    2.12&   -0.11&   -0.61&   -0.61\\
    -0.39&    0.01&   -0.11&    1.88&    0.49&    0.46\\
    -0.26&    0.56&   -0.61&    0.49&    2.37&    0.20\\
    -0.25&    0.05&   -0.61&    0.46&    0.20&    1.24\\
    \end{bmatrix},\\
    R_i&=0.3818,\;\;\forall i\in\mathcal{V}.
    \end{aligned}
    $$
    
    The explicit value of $\mathcal{L}$ is determined with $l_{ij}=\frac{a_{ij}}{d_{ii}}$, where $a_{ij}=1$ if $(i,j)\in\mathcal{E}$, and $d_{ii}=\sum_{j=1}^{N}a_{ij}$. Hence the matrix $\mathcal{L}$ obtained in this way is row-stochastic and primitive.
    Through performing the iterative law \eqref{HRiter}, one can obtain the solution to the HCREs \eqref{HCRE}. 
    
\begin{figure}
    	\centering
    	\includegraphics[width=0.5\textwidth]{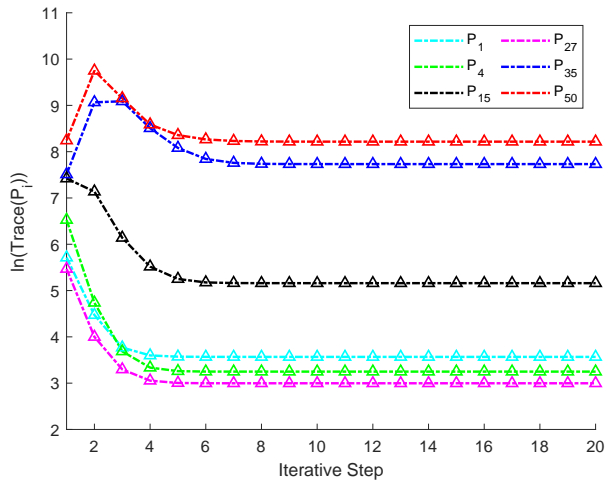}
    	\caption{
    		Illustration figure for the iteration of the trace of matrix $P_{i,k}$, where $P_{i}$ denotes the iterative value of $P_{i,k}$.}
    	\label{Iteration}	
    \end{figure}

\begin{figure}
	\centering
	\includegraphics[width=0.5\textwidth]{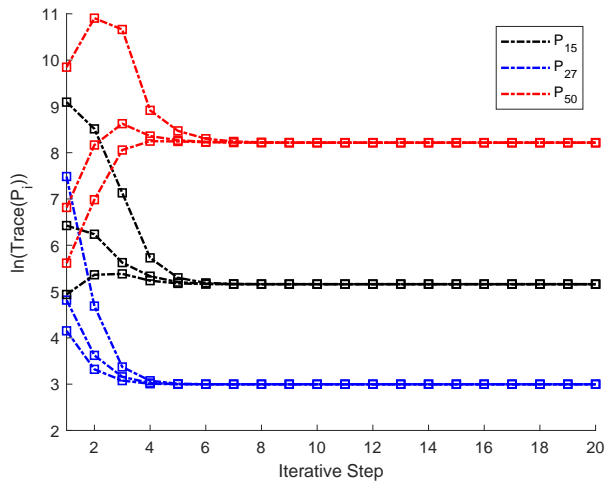}
	\caption{
		Illustration figure for the iteration of the trace of matrix $P_{i,k}$ with different initial value, where $P_{i}$ denotes the iterative value of $P_{i,k}$.}
	\label{Initial}	
\end{figure}

    In Fig.~\ref{Iteration}, six of the iterative values $P_{i,k}$ are presented, where $P_{i}$ denotes the iteration of the trace of matrix $P_{i,k}$. It is shown that each $P_{i,k}$ converges to the steady-state form, i.e., the solution to HCREs \eqref{HCRE}, through the iterative law \eqref{HRiter}.
    In Fig.~\ref{Initial}, three of the iterative values $P_{i,k}$ with 3 different initial values $P_{i,0}$ are presented. One can find that the convergence value $P_{i}$ is not related to the initial value, which verifies the uniqueness of the solution to HCREs.

   In addition, a target tracking numerical experiment is also provided to verify the theory proposed in Section \ref{secdiscussion} related to CIDF. The state transition matrix has the expression
   $$
   \begin{aligned}
   &a_{k}=\begin{pmatrix}
   1&T\\
   0&1
   \end{pmatrix},\qquad
   &A_k=\begin{pmatrix}
   a_{k}&0_{2\times 2}\\
   0_{2\times 2}&a_{k}
   \end{pmatrix}.
   \end{aligned}
   $$ 
   The error covariance matrix $Q$ {takes the form of}
   $$
   \begin{aligned}
   &G=\begin{pmatrix}
   \frac{T^3}{3}&\frac{T^2}{2}\\
   \frac{T^2}{2}&T
   \end{pmatrix},\quad
   &Q=\begin{pmatrix}
   G&0.5G\\
   0.5G&G\\
   \end{pmatrix},
   \end{aligned}
   $$
   where the sample interval is set to be $T=1$. The observation models of the three kinds of sensors are modified as:
   $$
   \begin{aligned}
   &C^{(1)}=\left[1,0,0,0\right],\\
   &C^{(2)}=\left[0,0,1,0\right],\\
   &C^{(3)}=\left[0,0,0,0\right].
   \end{aligned}
   $$
   The strucure of the network remains to be the same as that of the first numerical experiment, including 3 sensors of kind $C^{(1)}$, 3 sensors of kind $C^{(2)}$, 44 sensors of kind $C^{(3)}$, and $R_{i}=1,\forall i\in\mathcal{V}$. 
   
   Using Monte Carlo method, the filtering process is run 100 steps for each simulation and 1000 times in total. In this experiment, the performance of the CIDF algorithm is evaluated by the mean square error (MSE), i.e.,
   \begin{equation}\label{mse}
   \text{MSE}_{i,k}=\frac{1}{1000}\sum_{l=1}^{1000}\big\|\hat{x}_{i,k}^{\left(l\right)}-x_{k}^{(l)}\big\|_{2}^{2},
   \end{equation}
   and the mean of the MSE of all sensors at time step $k$ takes the expression as 
   \begin{equation}\label{mymse}
   \text{MSE}_k=\frac{1}{N}\sum_{i=1}^{N}\text{MSE}_{i,k},
   \end{equation}
   where $\hat{x}_{i,k}^{\left(l\right)}$ and $x_{k}^{(l)}$ denote the estimated state and real state at time step $k$ in the $l$-th simulation, respectively. 
   With the result {obtained} in Section \ref{secdiscussion}.A, in each simulation, the estimation error covariance matrix of the CIDF algorithm converges to the steady-state performance, which can be simplified as the solution $\mathcal{P}$ to DLE \eqref{DLE}. Hence, one can compare the trace of $\frac{1}{N}\mathcal{P}$ and $\text{MSE}_k$ to verify the theory proposed in Section \ref{secdiscussion}.A.
   \begin{figure}
   	\centering
   	\includegraphics[width=0.5\textwidth]{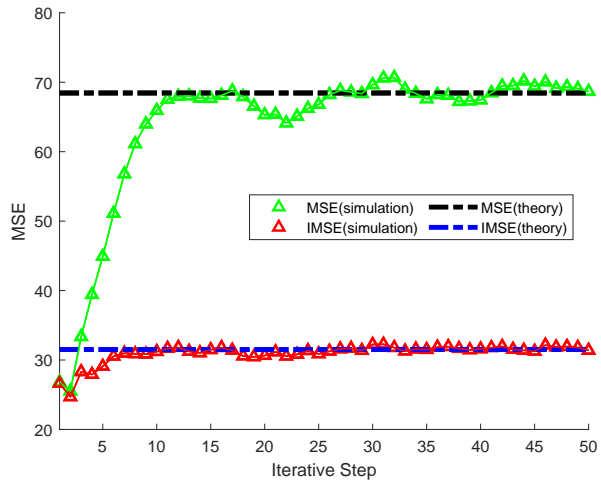}
   	\caption{
   		Illustration figure for the comparison between CIDF proposed in \cite{battistelli2014kullback} and ICF proposed in \cite{kamal2013information}, where MSE (simulation) and IMSE (simulation) denote the average MSE of the CIDF and ICF computed through xx, MSE (theory) and IMSE (theory) denote the trace of solution $\mathcal{P}$ to the DLE \eqref{DLE}.}
   	\label{meanSquare}	
   \end{figure}

   In Fig.~\ref{meanSquare}, the explicit MSE computed through \eqref{mymse} and theoretical MSE, i.e., trace of $\mathcal{P}$ are provided, where different CI-based distribtued filtering algorithms are also considered. The performance of CIDF proposed in \cite{battistelli2014kullback} and that of the ICF proposed in \cite{kamal2013information} are mainly compared with numerical experiment. From Fig.~\ref{meanSquare}, one can find that the mean square error will converge to the steady-state performance with $k\to\infty$, which verifies the correctness of the theory proposed in Section \ref{secdiscussion}.A. Moreover, as shown in Fig.~\ref{meanSquare}, the performance of the ICF is better than CIDF when the observable nodes are sparsely located in the sensor network, which indicates that one can optimize the performance of CIDF algorithm through locally tuning the parameter $\omega_{i,k}$. Note that the closed-form of the performance of CIDF algorithm is formulated as the solution to DLE \eqref{DLE}, future work may contain a further invstigation of the connection between the gain $\omega_{i,k}$ and the property of the solution $\mathcal{P}$ to DLE \eqref{DLE} to obtain efficient parameter optimization techniques.   
   
\section{Conclusion}\label{secconcu}
\textcolor{black}{In this paper, we have investigated the properties of the solution to a newly formulated harmonic-coupled Riccati equations (HCREs). We have shown that the uniqueness of the solution to HCREs can be guaranteed with the collective observability and primitiveness of the weighting matrix $\mathcal{L}$. Additionally, we have demonstrated that the matrix iterative law proposed in the CI-based distributed filtering (CIDF) algorithm converges to the solution to HCREs as the iteration step tends to infinity. Leveraging the newly discovered properties, we have simplified the closed-form of the steady-state estimation error covariance matrix of the CI-based distributed filtering algorithms as the solution to a discrete-time Lyapunov equation (DLE), the parameters of which are determined by solving the corresponding HCREs. Moreover, we have shown that the performance analysis of some well-known CI-based distributed filtering algorithms can also be unified under the framework of HCREs.}

	Future research will focus on investigating the relationship between the weighting parameters and the solution to HCREs and developing effective parameter tuning techniques to control the performance of the CIDF algorithm.

\appendices

\textcolor{black}{\bibliographystyle{IEEEtran}
\bibliography{ref}}

\end{document}